\documentclass[12pt]{iopart}

\usepackage{graphicx}  
\usepackage{amssymb}

\begin{document}

\title{Gamma Ray Spectroscopy with Scintillation Light in Liquid Xenon}

\author{K. Ni\footnote{Present address: Physics Department, Yale University, New Haven, CT 06511, USA.}, E. Aprile, K.L. Giboni, P. Majewski\footnote{Present address: Department of Physics and Astronomy, University of Sheffield, UK.}, M.~Yamashita} 
\address{Physics Department and Columbia Astrophysics Laboratory\\Columbia University, New York, New York 10027, USA}
\ead{kaixuan.ni@yale.edu}

\begin{abstract}
Scintillation light from gamma ray irradiation in liquid xenon is detected by two Hamamatsu R9288 photomultiplier tubes (PMTs) immersed in the liquid. UV light reflector material, PTFE, is used to optimize the light collection efficiency. The detector gives a high light yield of 6 photoelectron per keV (pe/keV), which allows efficient detection of the 122 keV $\gamma$-ray line from $^{57}$Co, with a measured energy resolution of $(8.8\pm0.6)$\% (\(\sigma\)). The best achievable energy resolution, by removing the instrumental fluctuations, from liquid xenon scintillation light is estimated to be around 6-8\% ($\sigma$) for $\gamma$-ray with energy between 662 keV and 122 keV. 
\end{abstract}


\section{Introduction}

Liquid xenon (LXe) has been used since many years ago for its good
ionization properties \cite{Carugno:NIM96,Aprile:LXeGRIT}, while applications based on its scintillation light are not fully developed until recently mostly due to inefficient detection of the UV scintillation light at 175 nm. Some of the recent applications, based on the LXe scintillation light, are using photomultiplier
tubes (PMTs) in the liquid (e.g. MEG experiment \cite{MEG} and the ZEPLIN-III detector \cite{ZEPLINIII}) or optical reflectors (e.g. XMASS experiment \cite{Yamashita:04} and the ZEPLIN-II detector \cite{ZEPLINII}) to achieve high light collection efficiency. With the improved light collection, better energy resolution and lower energy threshold can be achieved. Our group started to develop a new system with PMTs immersed in the liquid and with PTFE UV light reflectors to increase the light collection efficiency \cite{Aprile:IEEE03}, as part of the R\&D program for the XENON dark matter search \cite{XENON}. In this paper, we report the studies of light yield and energy resolution from LXe scintillation light for different $\gamma$-ray energies, from a small detector with optimized light collection efficiency. 

\section{Experimental setup}
\begin{figure}
\centering
\includegraphics[width=0.65\textwidth]{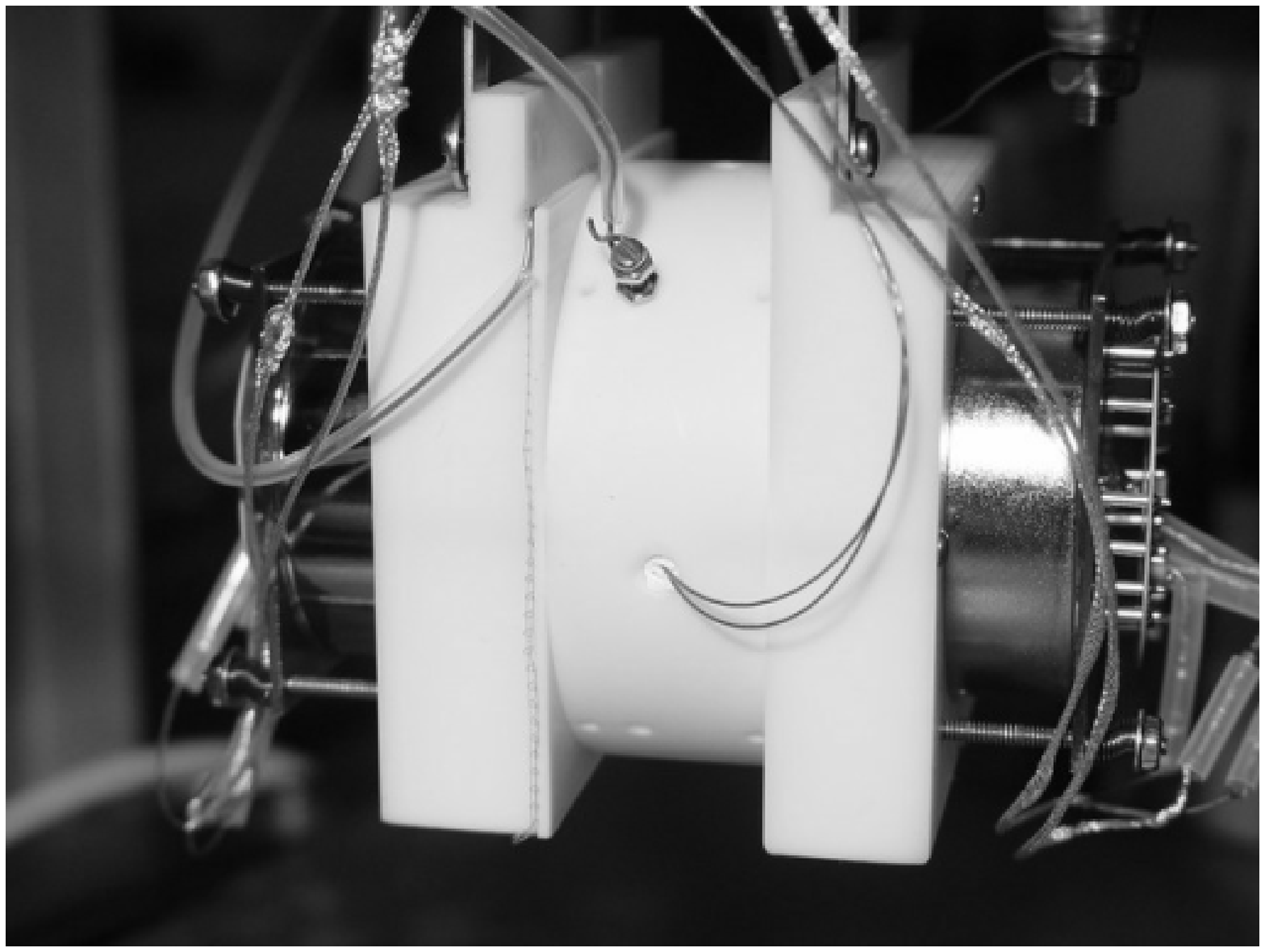}
\includegraphics[width=0.8\textwidth]{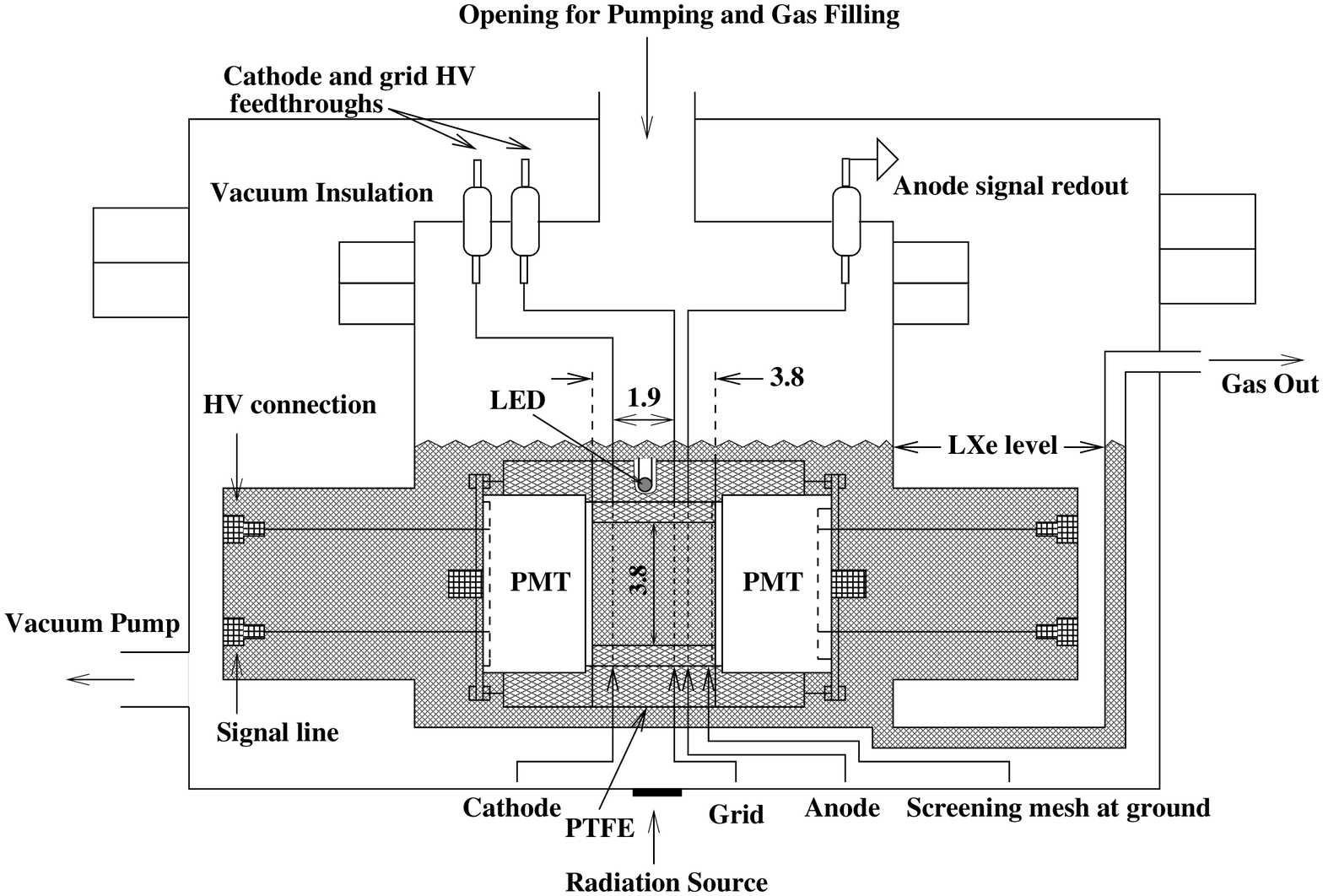}
\caption{A picture of the chamber inner structure (top) and a schematic drawing of the chamber (bottom, unit: cm).}
\label{chamber}
\end{figure}

A picture of the detector's inner structure and a schematic drawing are shown in Figure~\ref{chamber}. The detector contains a gridded ionization structure, enclosed by a PTFE tube. It was designed for both ionization and scintillation measurements \cite{Aprile:QL}. Two Hamamatsu R9288 PMTs are installed from the sides to detect the scintillation light. A blue LED is mounted in the PTFE tube. The light from LED goes through the PTFE wall and is detected by the PMTs for gain calibration. For ionization measurement, three high optical transparent meshes, which serve as cathode, grid and anode, are used. The ionization electrons are detected by the anode mesh, which is connected to a charge sensitive preamplifier (CLEAR PULSE Model 580). One additional shielding mesh is placed between the PMT and anode mesh to avoid induction between the PMT and the andoe charge signals. The 1.9 cm drift region between cathode and grid defines the liquid xenon sensitive volume for ionization measurement. For measurements of  only scintillation light, as in this paper, the whole volume of LXe between the two PMTs are the sensitive target. The whole structure is immersed in liquid xenon during the experiment. The structure is mounted in a stainless steel vessel, surrounded by a vacuum cryostat for thermal insulation. 

The detector is filled with ultra-pure liquid xenon, purified through a SAES high
temperature getter. After filling the chamber with liquid xenon, the
xenon is continuously circulated through the purification system, designed for the XENON dark matter search experiment (see \cite{ucla04} and \cite{dark04} for more details), to achieve the best xenon purity
level. During the recirculation of xenon, the light yield were monitored as a function of recirculation time and we saw no significant changes of light yield. This indicates that the
attenuation of scintillation light in liquid xenon is much longer than the size of our chamber. 

\begin{figure}
\centering
\includegraphics[width=0.9\textwidth]{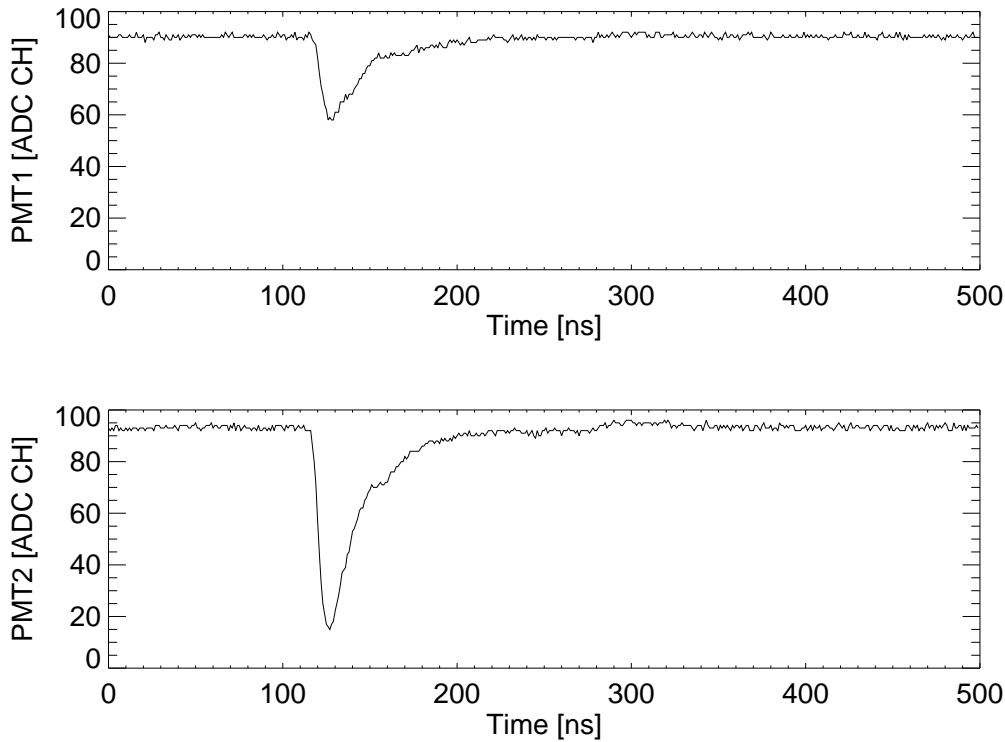}
\caption{Scintillation light waveforms from a typical 662 keV $\gamma$-ray event in LXe. The different sizes from the two PMTs indicate that the event location is closer to PMT2.}
\label{signal_wave}
\end{figure}

The whole scintillation light waveforms from the two PMTs are recorded by a digital oscilloscope (LeCroy model LT374) with a sampling rate of 1 GHz. 
The coincidence of the two signals is used for 
the trigger. Figure \ref{signal_wave} shows the two PMTs' output of the scintillation light waveforms from a 662 keV \(\gamma\)-ray ($^{137}$Cs) event. The integrated pulse area of the light waveforms gives the number of
photoelectrons ($N_{pe}$) produced by the scintillation light on the PMT photocathode, 

\begin{equation}
\label{eq:n_pe}
N_{pe} = \frac{\int{V(t)dt}/R}{e \cdot g}.
\end{equation}

Where $V(t)$ is the pulse amplitude at time $t$. $R$, equal to 50 $\Omega$, is the
impedance for the PMT signals. $e$ is the elementary electronic charge
and $g$ is the PMT gain. The gain of PMT is calibrated by the
single photoelectron spectrum from the LED light. 
The typical gain for the PMT at LXe temperature (-90$\rm ^o$C to -100$\rm ^o$C) is at the level of $10^6$ at an operating voltage of 900 V. 

\section{Simulation of light collection efficiency}
A Monte Carlo simulation, based on Geant4 \cite{geant4}, is developed to trace the scintillation photons produced in liquid xenon. In the simulation, we assume a PTFE reflectivity between 88\%$\sim$95\% \cite{Yamashita:04}. The optical transparency of the meshes is $95 \pm 1\%$. The absorption length of liquid xenon scintillation light is longer than 100 cm \cite{Baldini:04} and the Rayleigh scattering length is about 30 cm \cite{Ishida:97,Seidel:02}. The absorption and scattering lengths are much longer than the scale of our detector. The major uncertainty for calculating the light collection efficiency is from the PTFE reflectivity and mesh transparency. By varying the PTFE reflectivity between 88\% and 95\% and mesh transparency between 94\% and 96\% in the simulation, an overall
light collection efficiency is estimated to be $\rm (63\pm 4)\%$ for events uniformly distributed in the LXe sensitive volume.  

\begin{figure}
\centering
\includegraphics[width=0.9\textwidth]{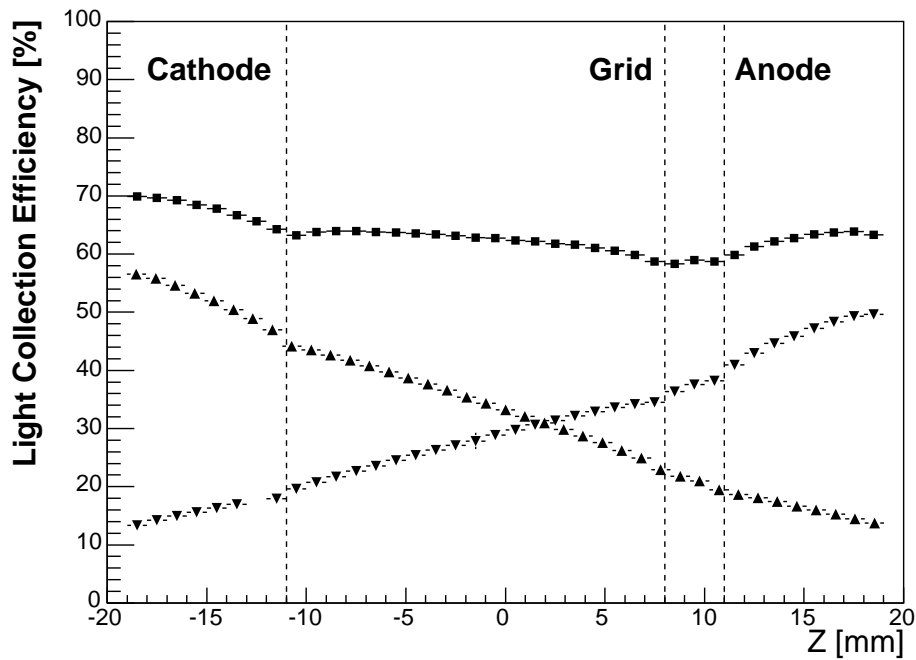}
\caption{Dependence of light collection efficiency with event location in $Z$. The downward triangles are for the right PMT and upward triangles are for the left PMT. The square is the sum from two PMTs. The three vertical dashed lines indicate the locations of cathode, grid, anode meshes, accordingly.} 
\label{lce_all}
\end{figure}

The solid angle effect makes the light detection sensitive to the event location. Figure \ref{lce_all} shows the position dependence of the simulated light
collection with PTFE reflectivity at 92\% and mesh transparency at 95\%. The horizontal center of the detector is at $Z = 0$. The surface of the two PMTs are at $\pm$19~mm. The light collection efficiency for each of the PMTs alone varies from 14\% to 60\% at different $Z$ positions. Combining the two PMTs' signals significantly reduces the non-uniformity. The asymmetry of the two PMT signals can be used to select the events in the center of the detector.

\section{Experimental Results}
\subsection{Scintillation yield}
\begin{figure}
\centering
\includegraphics[width=0.8\textwidth]{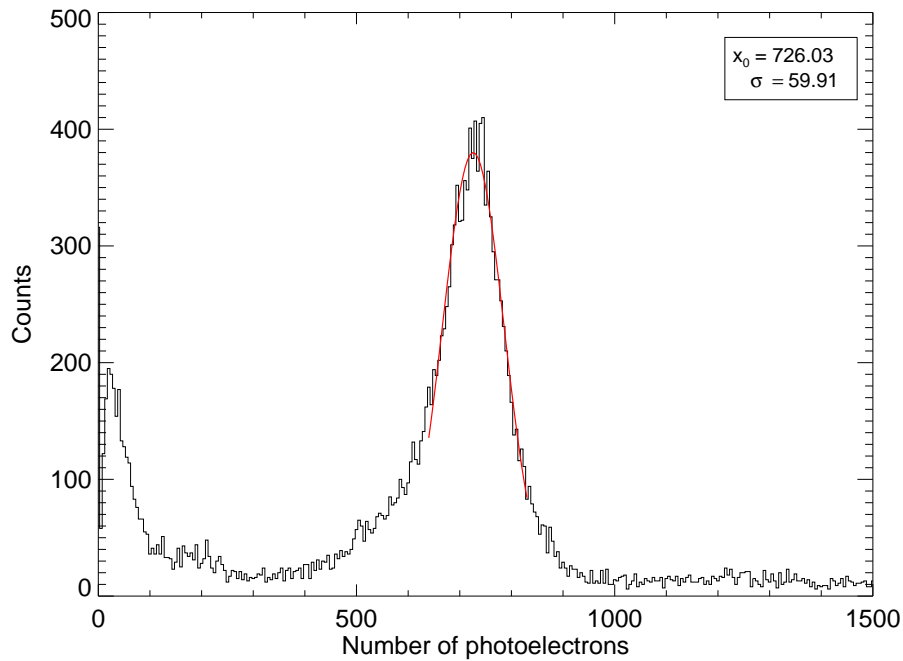}
\caption{$^{57}$Co scintillation light spectrum at zero field. A
  Gaussian fitting of the 122 keV peak gives a light yield of 726 pe. } 
\label{g4:co}
\end{figure}

\begin{figure}
\centering
\includegraphics[width=0.8\textwidth]{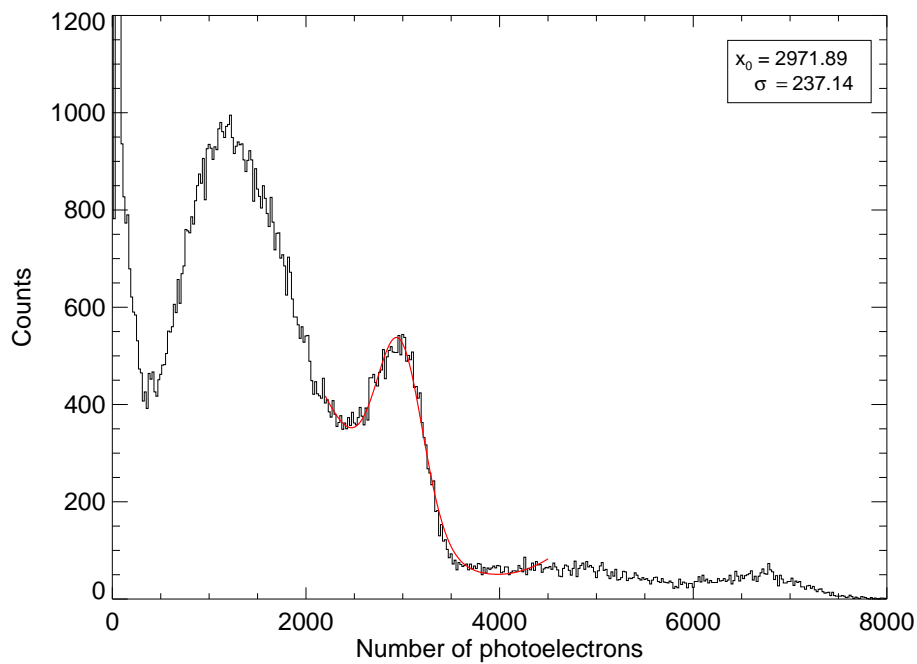}
\caption{\(^{22}\)Na scintillation light spectrum at zero field. The 511 keV peak is fitted with a Gaussian plus a second order polynomial function.}
\label{na_spe}
\end{figure}

\begin{figure}
\centering
\includegraphics[width=0.8\textwidth]{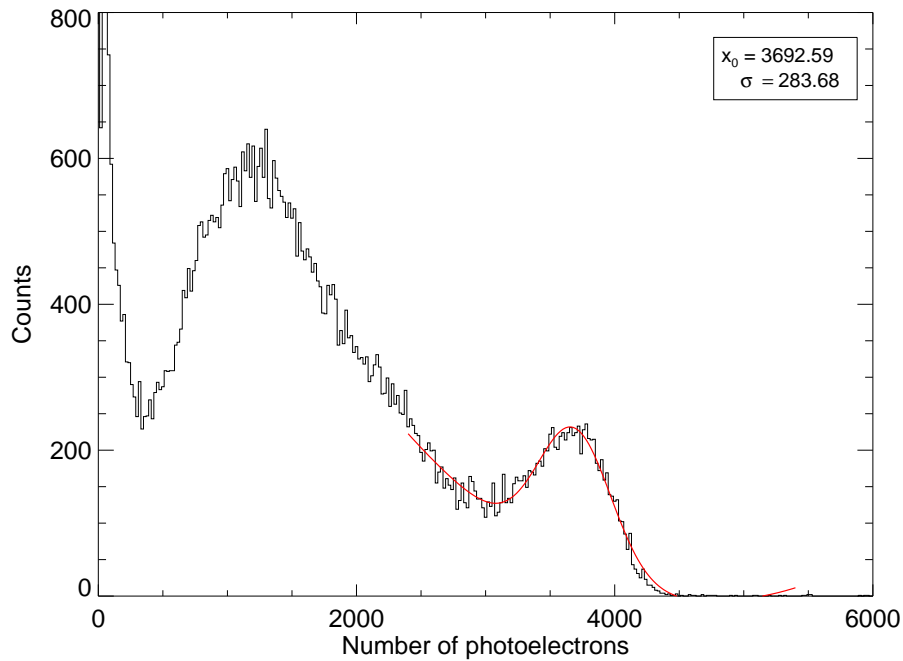}
\caption{\(^{137}\)Cs scintillation light spectrum at zero field. The 662 keV peak is fitted with a Gaussian plus a second order polynomial function. }
\label{cs_spe}
\end{figure}

The number of photoelectrons, $N_{pe}$, from a $\gamma$-ray event with energy $\varepsilon$ (in unit of eV) deposited in LXe can be calculated as, 
\begin{equation}
\label{eq:w_ph}
N_{pe} = \frac{\eta \cdot \varepsilon}{W_{ph}}\times L_c \times Q_e 
\end{equation}
where scintillation efficiency $\eta$ is defined as the gamma ray scintillation yield per unit energy compared with that of relativistic heavy ions. It is dependent on the $\gamma$ ray energy. $W_{ph}$ is the average energy required to produce a scintillation photon from the relativistic heavy ions, which is measured to be 13.8 eV \cite{Doke:01}. $L_c$ is the light collection efficiency. $Q_e$ is the effective quantum efficiency of the PMT (including photoelectron collection efficiency to the first dynode). $Q_e$'s for the PMTs used here have nominal values of 15\%. Figure~\ref{g4:co} is the $^{57}$Co scintillation light spectrum in number of photoelectrons, at zero field. The 122 keV $\gamma$ ray scintillation in LXe gives 726 pe, which corresponds to about 6.0 pe/keV. Using the simulated light collection efficiency, we estimate the $\eta$ value to be $0.85\pm0.10$ for the 122 keV $\gamma$ rays, according to equation \ref{eq:w_ph}. The error comes from the light
collection efficiency and PMT quantum efficiency. This measurement is repeated for $^{22}$Na (Figure~\ref{na_spe}) and $^{137}$Cs (Figure~\ref{cs_spe}) gamma ray sources. The light yields, estimated scintillation efficiencies and the energy resolutions are listed in Table 1.

\begin{center}
\begin{tabular}{ccccc}
\multicolumn{5}{c}{Table 1: Light yield and energy resolution } \\ 
\hline
Source & Energy  & Light yield & Scintillation efficiency & Energy resolution \\
 & [keV] & [pe/keV] & $\eta$ & [$\sigma$/E] \\
\hline
\(^{57}\)Co & 122 & 5.95 &  $0.85 \pm 0.10$  & $8.8 \pm 0.6$\% \\
\(^{22}\)Na & 511 & 5.82 &  $0.83 \pm 0.10$  & $8.0 \pm 0.3$\% \\
\(^{137}\)Cs & 662 & 5.58 & $0.79 \pm 0.09$  & $7.9 \pm 0.3$\% \\
\hline
\end{tabular}
\label{table2}
\end{center}

\begin{figure}
\centering
\includegraphics[width=0.95\textwidth]{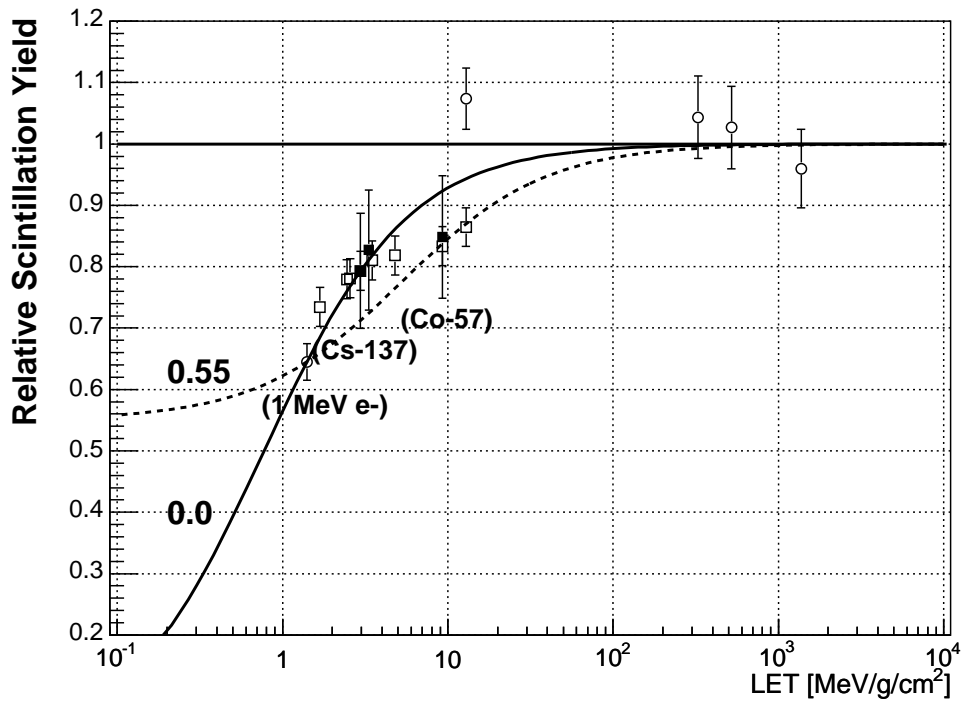}
\caption{LET dependence of scintillation light yield in liquid xenon
  for different energy $\gamma$ rays (open squares from
  \protect\cite{Yamashita:04} and solid squares from this paper), electrons
  and relativistic heavy ions (open circles from \protect\cite{Doke:01}). The
  LET values are obtained from \protect\cite{NIST}. The solid and dashed lines represent two different values (0 and 0.55) of $\eta_0$ for equation \ref{eq:let}.} 
\label{let}
\end{figure}

The measured energy dependence of scintillation light yield for $\gamma$
rays is consistent with the results reported in
\cite{Yamashita:04}, as shown in Figure~\ref{let}. The scintillation efficiency $\eta$ is usually referred as the relative scintillation yield per unit energy, $(dL/d\varepsilon)_{rel}$. $(dL/d\varepsilon)_{rel}$ is higher for $\gamma$ rays with lower energy. Its dependence on the linear energy transfer (LET), $d\varepsilon/dx$, is given by the following equation \cite{Doke:01},  
\begin{equation}
\label{eq:let}
(dL/d\varepsilon)_{rel} = \frac{A(d\varepsilon/dx)}{1+B(d\varepsilon/dx)} + \eta_0
\end{equation}
$A$, $B$ and $\eta_0$
are free parameters. $\eta_0$ is the $(dL/d\varepsilon)_{rel}$ value at zero LET. At $d\varepsilon/dx \rightarrow \infty$, $A/B + \eta_0 = 1$. The
relation between $A$ and $\eta_0$, and $B$ and $\eta_0$, can be
calculated with experimental values from 1 MeV electrons, which are
well precisely measured (less than 7\% uncertainty, as discussed in \cite{Doke:01}). We fit the experimental values, by combining
our results, with those from \cite{Yamashita:04} and
\cite{Doke:01}. The best fit gives an $\eta_0$ value at 0.55 for low
energy gamma rays, such as 122 keV $\gamma$ rays from $^{57}$Co. While for gamma rays more than 500 keV, such as 662 keV $\gamma$ rays from $^{137}$Cs, the fitted parameter $\eta_0$ prefers a zero value. 

\subsection{Energy resolution}
The energy resolution of the scintillation spectra is studied for different $\gamma$ ray energies. The values are also listed in Table 1. The overall measured energy resolution $R$ of a liquid xenon scintillation detector is a combination of light collection variation due to the detector geometry $R_g$, the statistical fluctuation of number of photoelectrons from the PMTs $R_s$, the fluctuation of electron-ion recombination due to escape electrons $R_r$, and the intrinsic resolution from liquid xenon scintillation light $R_i$. Here the intrinsic energy resolution $R_i$ is due to the non-proportionality of scintillation yield due to secondary electrons, as discussed in \cite{Doke:Xenon01} and references therein. $R$ can be written as the following equation, assuming all these factors are not correlated.

\begin{equation}
\label{eq:res}
R^2 = R_g^2 + R_s^2 + R_r^2 + R_i^2
\end{equation}

Based on light collection simulation, $R_g$ for the 511 keV $\gamma$
from $^{22}$Na and 662~keV $\gamma$ from $^{137}$Cs are about 4.7\%. Due to a much localized interaction positions, $R_g$ is about 2.5\% for 122
keV $\gamma$ from $^{57}$Co. $R_s$ can be calculated roughly as $R_s =
1.1/\sqrt{N_{pe}}$, which includes the statistical fluctuations of number of photoelectrons and PMT gain variations. The intrinsic energy resolution, $R_i$, is estimated to be about 4\% (FWHM) for gamma ray energy between 0.1 and 2~MeV (see Figure 7 in \cite{Doke:Xenon01}). 

\begin{figure}
\centering
\includegraphics[width=0.9\textwidth]{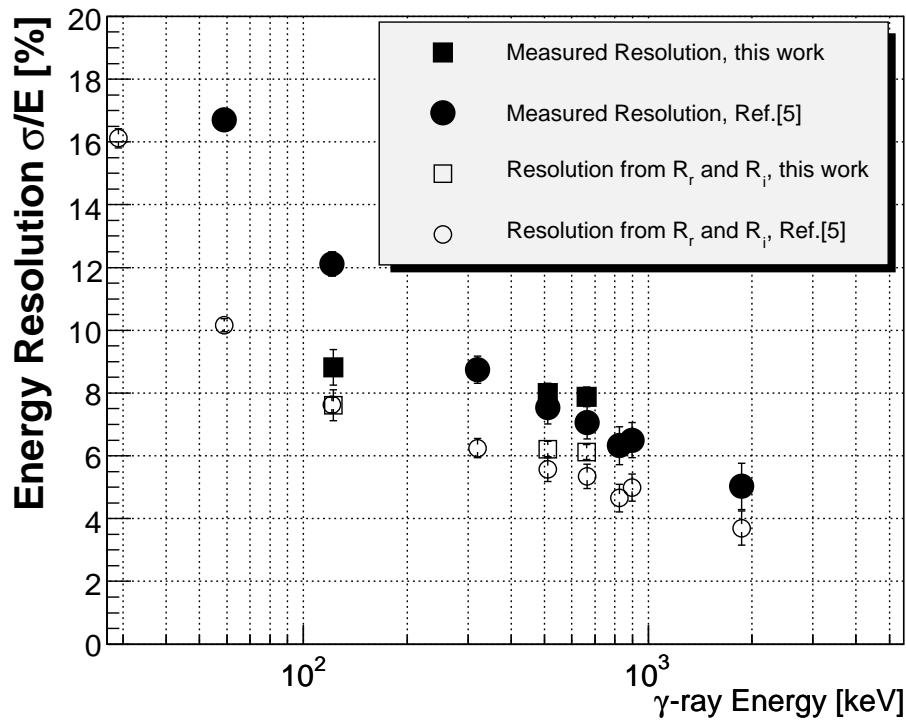}
\caption{Energy resolution dependence on the $\gamma$ ray energy in liquid xenon. The measured energy resolution (solid markers) can be further improved by reducing the statistical  and geometrical fluctuations of light collection. The resolutions contributed from the recombination fluctuation ($R_r$) and intrinsic resolution ($R_i$) (open markers) represent the best achievable resolutions from liquid xenon scintillation light alone.} 
\label{ene_res} 
\end{figure}

We can estimate the combined contribution from recombination fluctuation contribution $R_r$ and the intrinsic energy resolution $R_i$, based on the values of measured energy resolution $R$, the geometrical contribution $R_g$ and the statistical contribution $R_s$, from equation \ref{eq:res}. The results are plotted in Figure \ref{ene_res}, compared with the results in \cite{Yamashita:04}.

It is clear that energy resolution from liquid xenon scintillation light is limited by the recombination fluctuation $R_r$. At low energies, the statistical contribution $R_s$ is dominant.  Due to a better light collection efficiency, the statistical contribution is smaller in this work than that in \cite{Yamashita:04}. This explains the better measured energy resolution for 122 keV $\gamma$ rays in this work. The best achievable energy resolution, by using liquid xenon scintillation light at zero field, is about 6-8\%($\sigma$) for $\gamma$ rays with energy between 662 keV and 122 keV. We note that the recombination fluctuation can be removed technically by means of ionization-scintillation anti-correlation, and much better energy resolutions can be achieved in liquid xenon by using both scintillation and ionization signals, as from recent measurements in \cite{Aprile:QL,Conti:03} and discussions in \cite{Doke:05}.

\section{Conclusion}
In this paper we presented the results related to scintillation spectroscopy obtained from a liquid xenon chamber with a good light collection efficiency by immersing PMTs in the liquid and using UV light reflectors. The relative scintillation yield from gamma rays to relativistic heavy ions in liquid xenon is measured. The LET dependence of relative scintillation yield from $\gamma$ rays is discussed. The high light collection efficiency yields a good energy resolution of 8.8\%($\sigma$) for 122 keV $\gamma$ rays from $^{57}$Co at zero field. At zero field, the best energy resolution (i.e. without instrumental fluctuations) that can be achieved, by using liquid xenon scintillation light alone, is estimated to be around $6-8\%$($\sigma$) for $\gamma$ rays between 662 keV and 122 keV. 

\section{Acknowledgments}
This work is supported by a grant
from the National Science Foundation to the Columbia Astrophysics
Laboratory (Grant No.  PHY-02-01740) for the development of the XENON
Dark Matter Project. The authors would like to express their thanks to Tadayoshi Doke and Akira Hitachi for valuable discussions.

\end{document}